\documentclass[aps,
preprint,
groupedaddress,
amsmath,
amssymb,
]{revtex4-2}

\usepackage{graphicx}
\usepackage{dcolumn}
\usepackage{bm}

\begin{document}


\title{Spin orbit torque-driven motion of quasi-Bloch domain wall  in perpendicularly magnetized W/CoFeB/MgO structure}


\author{Nobuyuki Umetsu}
\affiliation{Frontier Technology R\&D institute, Kioxia Corporation, Yokohama, Japan.}
\author{Michael Quinsat}
\affiliation{Frontier Technology R\&D institute, Kioxia Corporation, Yokohama, Japan.}
\author{Susumu Hashimoto}
\affiliation{Frontier Technology R\&D institute, Kioxia Corporation, Yokohama, Japan.}
\author{Tsuyoshi Kondo}
\affiliation{Frontier Technology R\&D institute, Kioxia Corporation, Yokohama, Japan.}
\author{Masaki Kado}
\affiliation{Frontier Technology R\&D institute, Kioxia Corporation, Yokohama, Japan.}

\date{\today}

\begin{abstract}
The motion of chiral magnetic domain walls (DWs) driven by spin-orbit torque (SOT) has been extensively studied in heavy metal/ferromagnet heterostructures with perpendicular magnetic anisotropy.
This study specifically focuses on SOT-driven DWs in near Bloch-states, which we refer to as ``quasi-Bloch DWs". These quasi-Bloch DWs exhibit slower motion compared to Neel-type DWs, offering potential for achieving highly controllable DW positions.
Here, we investigate the characteristics of SOT-driven motion of quasi-Bloch DWs in perpendicularly magnetized ultra-thin films consisting of W/CoFeB/MgO.
For analyzing the DW motion, we employ a one-dimensional model incorporating parameters derived from experimental data obtained from our samples.
Our model successfully reproduces the experimental results, which reveal variations in the direction and threshold current density of DW motion among different samples. 
Through theoretical analysis, we unveil that the DW remains in quasi-Bloch states during motion, with SOT serving as the primary driving force rather than spin transfer torque (STT). 
The direction of motion is determined not only by the sign combination of Dzyaloshinskii-Moriya interaction (DMI) and spin Hall angle but also by the strength of DMI, STT, and extrinsic DW pinning. 
Furthermore, we provide analytical expressions for the threshold current density required for SOT-driven quasi-Bloch DW motion. These findings provide valuable insights for the design of future DW devices with specific film structures.

\end{abstract}


\maketitle


\section{Introduction}

The phenomenon of chiral magnetic domain wall (DW) motion driven by spin-orbit torque (SOT) has been extensively studied
in heavy metal (HM)/ferromagnet (FM) hetrostructures with perpendicular magnetic anisotropy (PMA)
\cite{Moore_2008, Kim_2010, Miron_2011, Haazen_2013, Emori_2013, Koyama_2013, Torrejon_2014, Ryu_2014, Torrejon_2016, Lau_2019, Yoon_2022}.
SOT offers the potential to drive DWs faster than conventional spin transfer torque (STT) \cite{Ryu_2013, Yang_2015, Kim_2017_ferri, Caretta_2018, Velez_2019, Ranjbar_2022}
and at lower currents \cite{Meng_2016, Thach_2024}.

The characteristics of SOT-driven DW motion are influenced by both the properties of SOT and the DW structure \cite{Thiaville_2012, Khvalkovskiy_2013},
which is governed by the Dzyaloshinskii-Moriya interaction (DMI) at the HM/FM interface.
In theory, SOT affects the DW motion except when it is in a perfect Bloch-state.
The strength of SOT acting on the DW is greater when the DW is closer to a N\'{e}el-state,
resulting in higher DW velocities. 
The direction of SOT-driven DW motion is determined by
the combination of the DW chirality and the sign of the spin Hall angle (SHA) \cite{Kim_2022}.
When the DW is close to N\'{e}el-states,
the chiral direction is determined by the sign of DMI,
but  when the DW is close to Bloch-states,
the chiral direction can also be influenced by STT and extrinsic DW pinning.
However, previous studies have not provided a comprehensive interpretation of the experimental results regarding the direction of DW motion from this standpoint. 
Although the fast SOT-driven motion of N\'{e}el-type DWs is intriguing, it may not be suitable for precise control of DW positions in integrated circuit devices.
This is because the sensitivity of DW travel distance to pulse width and pulse rise (fall) increases with higher velocity. Additionally, controlling the position of DWs becomes more challenging when using short pulses due to the significant RC delay in high-density devices with large memory cell arrays. Conversely, the SOT-driven motion of DWs in near Bloch-states, which we refer to as "quasi-Bloch DWs" \cite{Pamyatnykh_2001, Popov_2016, quasiBloch}, offers slower motion that is better suited for specific practical applications.

In this study, we investigate the characteristics of  SOT-driven motion of quasi-Bloch DWs
in perpendicularly magnetized ultra-thin films consisting of W/CoFeB/MgO.
We experimentally examine the magnetic film properties, efficiencies of SOT, and current-induced DW motion (CIDWM).
By utilizing these experimental values, 
we perform model calculations that incorporate not only SOT and DMI but also STT and the depinning magnetic field \cite{Jeudy_2018, Jagt_2022} as a substitute for extrinsic DW pinning.
Our model successfully reproduces the experimental results, which reveal variations in the direction and threshold current density of DW motion among different samples. 
Our theoretical analysis reveals that the DW remains in near Bloch-states during SOT-driven motion.
Additionally, we identify the key factors influencing the direction of DW motion and provide analytical expressions for the threshold current density of DW motion.
These findings are of utmost importance for the design of future DW devices' film structures \cite{Parkin_2015}.

The remainder of this paper is organized as follows:
In Sec. II, we present the characterization of our samples based on the measurement results of magnetic film properties, SOT efficiencies, and CIDWM.
In Sec. III,  we present our model and demonstrate its capability to reproduce the experimental results, which reveal variations in the direction and threshold current density of DW motion among different samples.
In Sec. IV, we discuss the direction, speed, and threshold current density of SOT-driven motion for quasi-Bloch DWs based on our model.
We specifically focus on the influence of STT, DMI, and extrinsic DW pinning on the motion characteristics.
Finally, we summarize our findings in Sec. V.

\section{Sample characterization}

In this study, we used ultra-thin films consisting of  Al$_2$O$_3$($20\,\mathrm{nm}$)/
W(2$\,\mathrm{nm}$)/
Co$_{40}$Fe$_{40}$B$_{20}$(1-1.3$\,\mathrm{nm}$)/MgO(2$\,\mathrm{nm}$)/TaN(1$\,\mathrm{nm}$).
The films were deposited onto thermally oxidized Si substrates
using  a magnetron sputter deposition system with a base pressure of less than $5 \times 10^{-6}\,\mathrm{Pa}$.
The deposition pressures for W and CoFeB layers were 0.04-0.5$\,\mathrm{Pa}$ and $0.27\,\mathrm{Pa}$, respectively.
The samples were annealed at $300\,^{\circ}\mathrm{C}$ or $400\,^{\circ}\mathrm{C}$ in a vacuum for 1 hour.
The thickness of the CoFeB layer $(t_{{\rm CFB}})$, the deposition pressure of the W layer $(P_{\rm W})$,
and the annealing temperature $(T_{{\rm a}})$ for each sample are summarized
in Table \ref{tab:table1}.

\begin{table}[htbp]

\centering
\begin{tabular}{cccc}
\hline
\textrm{ Sample }&
\textrm{ $t_{{\rm CFB}}$ ({\rm nm})}&
\textrm{ $P_{{\rm W}}$ ({\rm Pa})}&
\textrm{ $T_{{\rm a}}$ ($^{\circ}$C)}\\
\hline
A & 1.2 & 0.27 & 300\\
B & 1.3 & 0.5 & 300\\
C & 1.0 & 0.04 & 400\\
D & 1.1 & 0.04 & 400\\
E & 1.2 & 0.27 & 400\\
\hline
\end{tabular}

\caption{\label{tab:table1}%
Deposition conditions for each sample.
}

\end{table}

We evaluated various magnetic film properties and SOT efficiencies.
The saturation magnetization, $M_{{\rm s}}$ and the
effective PMA, $K_{{\rm eff}}$ were
measured using vibrating sample magnetometry (VSM).
The magnetization curves measured by VSM are shown in Appendix \ref{sec:VSM}.

The STT efficiency is estimated using the obtained values of $M_{{\rm s}}$ and $K_{{\rm eff}}$ with the following equation:
\begin{align}
\xi_{{\rm ST}}=-\frac{\hbar P}{2\mu_{0}\gamma\left|e\right|M_{{\rm s}}\lambda},
\end{align}
where, $\mu_{0}$ is the magnetic permeability, $\gamma$ is the gyromagnetic ratio, 
$e$ is the elementary charge, $\hbar$ is the Dirac constant, $P$ is the spin polarization, and
$\lambda=\sqrt{A/K_{{\rm eff}}}$ is the half width of the DW (with $A$ as the exchange stiffness constant).
For the estimation, we assume typical values of $A$ = 15$\,\mathrm{pJ/m}$ \cite{Jagt_2022, Kato_2019} and $P=0.5$.

We determined the depinning magnetic field, $H_{{\rm dep}}$, through magnetic bubble domain expansion experiments \cite{Pardo_2017, Jeudy_2018, Jagt_2022}.
The detailed method and measurement data are provided in Appendix \ref{sec:Depin}.
The value of $H_{{\rm dep}}$ is defined as the boundary value between the pinning-dependent thermally activated creep regime and the depinning transition regime.
Since $H_{{\rm dep}}$ corresponds to the threshold field for DW depinning at zero temperature \cite{Pardo_2017},
we consider this value as a substitute for the extrinsic DW pinning field in our model, as discussed in the next section.

The DMI field, $H_{{\rm DMI}}$ \cite{Rohart_2013} was also obtained through magnetic bubble domain expansion experiments \cite{Quinsat_2017, Soucaille_2016, Dohi_2019}.
The measurement data are provided in Appendix \ref{sec:DMI}.
The DMI constant is evaluated using
\begin{align}
D=\mu_{0}M_{{\rm s}}\lambda H_{{\rm DMI}}.
\end{align}

We determined the Slonczewski and field-like components of SOT efficiencies, $\xi_{{\rm SL}}$ and
$\xi_{{\rm FL}}$, using harmonic Hall voltage measurements \cite{Hayashi_2014, Li_2018, Zhu_2019}.
The samples were patterned into Hall bar devices with a channel size of $5\times 5\,\mu$m$^2$ by using conventional micro-fabrication methods.
The detailed method and measurement data are provided in Appendix \ref{sec:Harmonic}.
When estimating $\xi_{{\rm SL(FL)}}$,
it is necessary to account for the planar Hall effect (PHE) in addition to the anomalous Hall effect (AHE) if the Hall voltage includes both contributions \cite{Hayashi_2014}.
We assume a  PHE resistance to AHE resistance ratio of 0.3 \cite{Skowronski_2016}.
In the subsequent section, we demonstrate that the PHE contribution to the discussion in this study is negligibly small.

The spin Hall angle, $\Theta_{\rm SH}$ and the ratio of the field-like torque to the Slonczewski torque, $\beta_\mathrm{SO}$
are evaluated using  the experimental values of $\xi_{{\rm SL (FL)}}$ with the following equations:
\begin{align}
\xi_{{\rm SL}}=\frac{\hbar\Theta_{{\rm SH}}}{2\mu_{0}\gamma\left|e\right|M_{{\rm s}}t_{{\rm CFB}}},\quad
\xi_{{\rm FL}}=\beta_{\mathrm{SO}}\xi_{{\rm SL}}.
\end{align}
Here,  we assume that the same current density occurs in both the CoFeB and W layers.

\begin{figure}[htbp]
\centering
\includegraphics[scale=0.5]{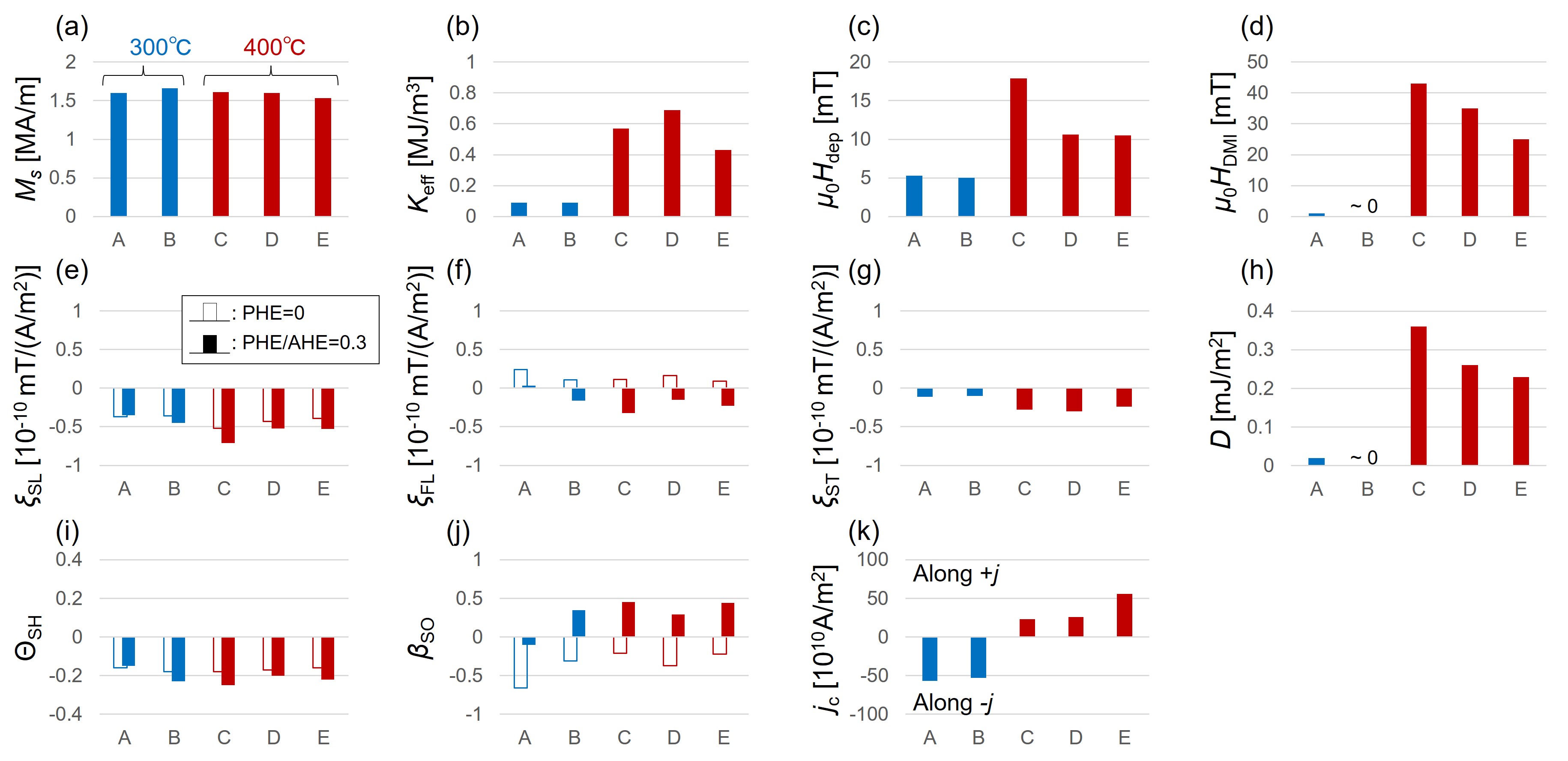}
\caption{\label{fig:properties} (a)-(j)  Magnetic film properties and parameters related to current-induced spin torques for each sample. The magnetic dead layer of $0.3\,\mathrm{nm}$ is taken into account in these estimations (See Appendix \ref{sec:VSM}).
(k) Threshold current density for DW motion.
}
\end{figure}

Figure \ref{fig:properties}(a)-(j) show the magnetic film properties and parameters related to current-induced spin torques for each sample.
The values of $K_{{\rm eff}}$, $\xi_{\rm ST}$, $H_{{\rm dep}}$, $H_{{\rm DMI}}$, and $D$
can be classified into two categories according to $T_{\rm a}$.
In W/CoFeB/MgO structures, it has been reported that annealing promotes the crystallization of W \cite{Hao_2015, Kim_2018} and the diffusion of B \cite{An_2015, Xu_2018}.
The higher value of DMI observed for $T_{\rm a}=400\,^{\circ}\mathrm{C}$ compared to $T_{\rm a}=300\,^{\circ}\mathrm{C}$ can be attributed to the enhanced crystallinity of the W/CoFeB interface. The $D\sim 0$ observed in the samples annealed at $300\,^{\circ}\mathrm{C}$ may indicate that the W layers are amorphous. The diffusion of B into the MgO layer affects the crystallization of CoFeB and the PMA at the CoFeB/MgO interface \cite{Cao_2014, An_2015}. The higher value of $K_{{\rm eff}}$ observed for $T_{\rm a}=400\,^{\circ}\mathrm{C}$ compared to $T_{\rm a}=300\,^{\circ}\mathrm{C}$ could be a result of the enhanced crystallinity of both the W/CoFeB and CoFeB/MgO interfaces. The samples annealed at $400\,^{\circ}\mathrm{C}$ exhibits higher values of both $K_{{\rm eff}}$ and DMI compared to those annealed at $300\,^{\circ}\mathrm{C}$, suggesting a potential correlation between interfacial PMA and DMI, as indicated by previous studies \cite{Khan_2016, Kim_2017, Cho_2018}.
The increase in $K_{\rm eff}$ leads to an increase in $\xi_{\rm ST}$ due to a decrease in the estimated value of $\lambda$. Additionally, the increase in $K_{\rm eff}$ may result in an increased spatial dispersion of $K_{\rm eff}$, which causes an increase in $H_{\rm dep}$ due to enhanced extrinsic pinning for DW motion \cite{Franken_2012, Balan_2023, Giuliano_2023, Tatara_2008}.

\begin{figure}[htbp]
\centering
\includegraphics[scale=0.45]{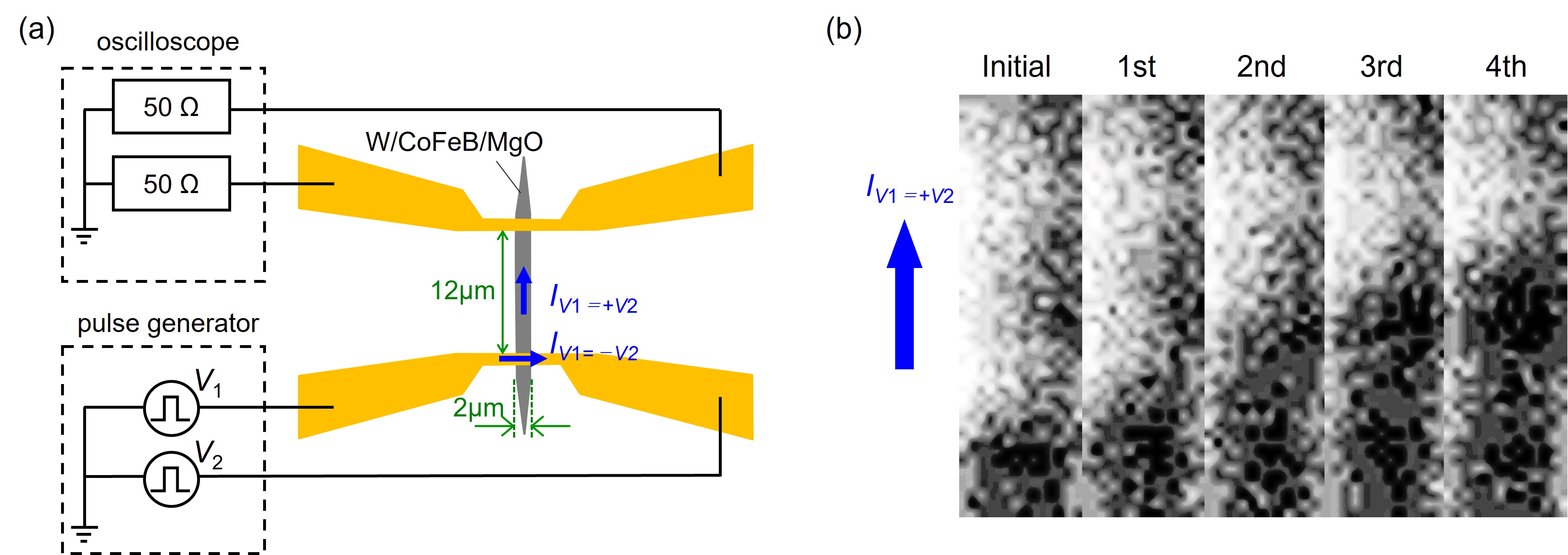}\caption{
\label{fig:CIDWM}
(a) Schematic illustration of setup for CIDWM experiments. (b) Examples of measured magnetic domain images.}
\end{figure}

We observed CIDWM and determined the threshold current density for DW motion for each sample.
The experimental setup is shown in Fig. \ref{fig:CIDWM}(a).
Nanowires with a width of $2\,\mu\mathrm{m}$ and Ta/Au electrodes in the form of a coplanar waveguide were fabricated \cite{Kado_2023}.
The samples were then processed into devices using conventional micro-fabrication methods.
The voltage pulses, $V_1$ and $V_2$, responsible for inducing DW writing and shifting in the nanowire, are generated by a pulse generator with dual output.
The value of $V_2$ is set to $-V_1$ and $+V_1$ to write and shift the DW, respectively. To ensure proper operation,
coaxial cables and probes of similar specification are used between the pulse outputs and the device's contact pad.
Additionally, the time delay between $V_1$ and $V_2$ is adjusted within the pulse generator to prevent any leakage or spike current from flowing into the nanowire during the DW writing phase ($V_2=-V_1$).
The electrical current flowing in the nanowire is measured by monitoring the voltage at the opposite electrodes using an oscilloscope with a 50$\,\Omega$ input impedance.
Changes in the DW position were observed using a magneto-optical microscope when multiple pulses of $20\,\mathrm{ns}$ width were applied, as shown in Fig. \ref{fig:CIDWM}(b).

Figure \ref{fig:properties}(k) shows the threshold current density, $j_{{\rm c}}$ for DW motion.
The sign of $j_{{\rm c}}$ denotes the direction of DW motion.
We observe that DW moves in the direction of electron flow in the samples annealed at $300\,^{\circ}\mathrm{C}$
and in the direction of current flow in the samples annealed at $400\,^{\circ}\mathrm{C}$.
In Ta/CoFeB/MgO and Pt/CoFeB/MgO structures, both directions of DW motion have also been observed \cite{Emori_2013, Conte_2015, Karnad_2018}.
In situations where the DW is locked in near N\'{e}el-states by strong DMI ($D\gg0.1\,\mathrm{mJ/m}^2$),
the direction of DW motion is consistent with the sign combination of $D$ and SHA \cite{Kim_2022}.
However, our experimental results cannot be explained using the same logic because
$D$ is nearly 0 in the samples annealed at $300\,^{\circ}\mathrm{C}$.
In Sec. IV, we theoretically elucidate the mechanism of DW motion parallel to the electron flow in such situations
where the DW is in near Bloch-states.

\section{Model}

Assuming the uniformity of the DW cross section perpendicular to the direction of its motion,
the dynamics of the DW can be described by the one-dimensional model (1DM)
\cite{Tatara_2008,Martinez_2014,Rizinggard_2017}.
In this model, the DW motion is represented by the position ($X$) and the magnetization angle ($\phi$),
where $\phi=0$ and $\pi$ correspond to N\'{e}el-type DWs,
and $\phi=\pm \pi /2$ correspond to Bloch-type DWs.
These time-dependent valuables are governed by the following equations:
\begin{subequations}
\label{eq:1DMeq}
\begin{align}
\frac{{\rm d}\phi}{{\rm d}t} = &\frac{\gamma}{1+\alpha^{2}}
\left[ -{Q}\Gamma_{1}\left(\phi,X\right)-\alpha\Gamma_{2}\left(\phi,X\right) \right],\\
\frac{{\rm d}\left(X/\lambda\right)}{{\rm d}t} =  &\frac{\gamma}{1+\alpha^{2}}
\left[ -\alpha\Gamma_{1}\left(\phi,X\right)+{Q}\Gamma_{2}\left(\phi,X\right) \right],\\
\Gamma_{1}\left(\phi,X\right) = &{Q}\frac{\pi}{2}\xi_{{\rm SL}}j\cos\phi+H_{{\rm p}}\left(X\right),\\
\Gamma_{2}\left(\phi,X\right) = &{Q}\xi_{{\rm ST}}j+\frac{\pi}{2}{Q}H_{{\rm DMI}}\sin\phi
-\frac{\pi}{2}\xi_{{\rm FL}}j\cos\phi-\frac{1}{2}H_{{\rm a}}\sin2\phi,
\end{align}
\end{subequations}
where,  $\alpha$ is the Gilbert damping constant,
${Q}$ is the topological charge of the wall
(${Q}=+1$ for $\uparrow\downarrow$ wall and ${Q}=-1$ for $\downarrow\uparrow$ wall),
$j$ is the current density ($j>0$ is considered in this article),
$H_{{\rm a}}=\ln2M_{{\rm s}}t_{{\rm CFB}}/\left(\pi\lambda\right)$ is the shape anisotropy field, and
$H_{{\rm p}}\left(X\right)$ is the extrinsic DW pinning field arising from the $Q$-independent pinning potential, which depends on the DW position \cite{Tatara_2008}.
We use the value of $\alpha=0.03$, which was experimentally determined by broadband ferromagnetic resonance measurements \cite{Nakamura_2024} for  sample-E.
Assuming that the annealing temperature dependence of $\alpha$ is not significant \cite{Lattery_2018},
we apply this value to all samples in our analysis.

Although STT and extrinsic DW pinning are often neglected when focusing on the qualitative behavior of SOT-driven DW motion \cite{Thiaville_2012, Khvalkovskiy_2013},
this study considers these terms for quantitative discussion.
We assume a sinusoidal pinning potential \cite{Martinez_2011, Torrejon_2014, Conte_2015}
and express the extrinsic DW pinning field as
$H_{{\rm p}}\left(X\right)=H_{{\rm pin}}\sin\frac{2\pi X}{L_{{\rm pin}}}$,
where $H_{{\rm pin}}$ is the amplitude and $L_{{\rm pin}}$ is the periodic length.
Although direct measurement methods for
$H_{{\rm pin}}$ and $L_{{\rm pin}}$ have not been established, $H_{{\rm dep}}$ is used as a substitute for $H_{{\rm pin}}$.
Additionally, based on the correlation length reported in the previous study \cite{Jeudy_2018}, we assume $L_{{\rm pin}}=20\,\mathrm{nm}$ \cite{periodic}.

\begin{figure}[htbp]
\centering
\includegraphics[scale=0.8]{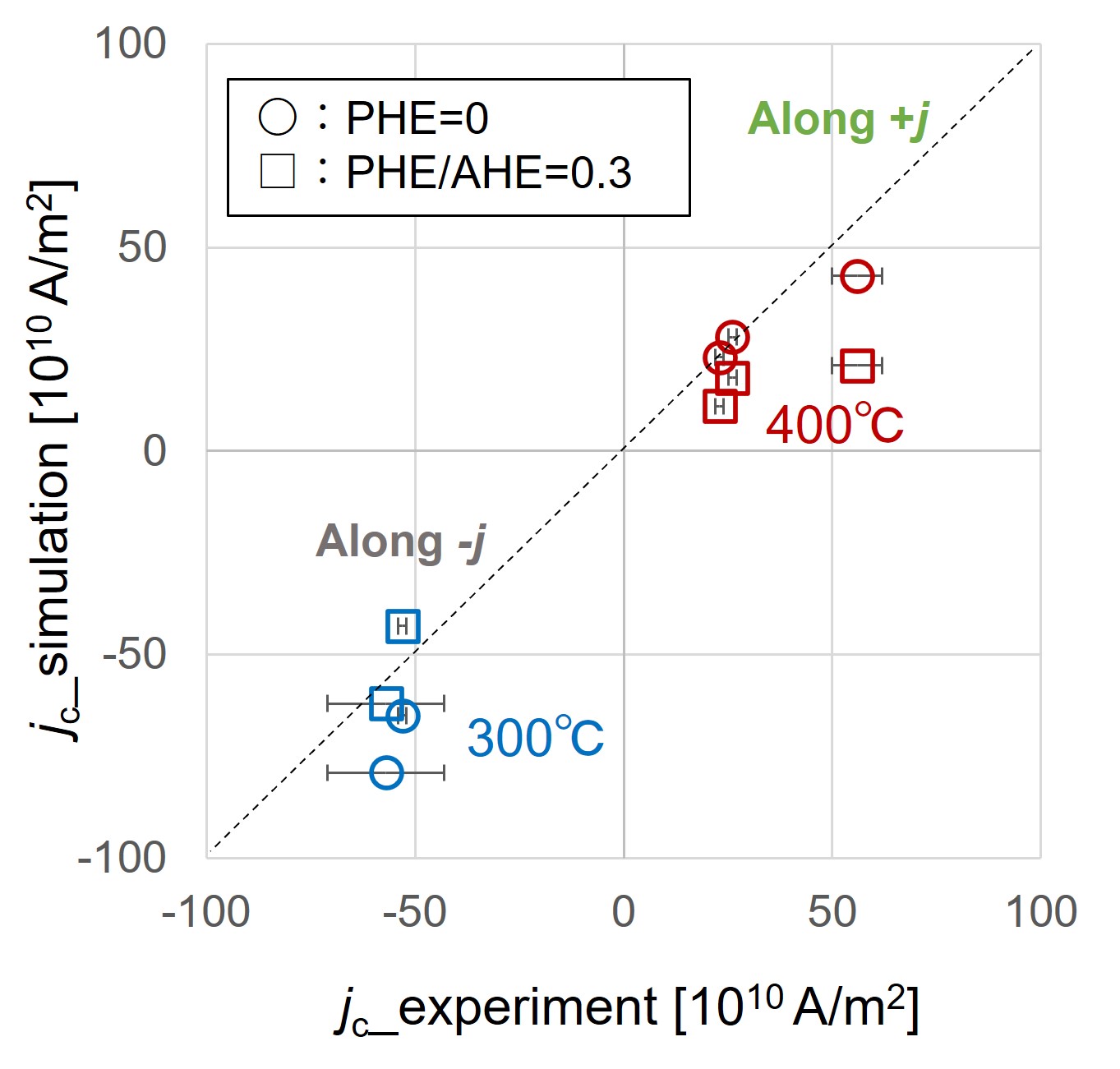}\caption{
\label{fig:sim_sample}
Comparison of experimental results with $j_{{\rm c}}$ obtained by 1DM simulation.
The measured results (Fig. \ref{fig:properties}(a)-(j)) were used as parameters for the simulation.
Error bars indicate the range of variability observed in multiple measurements, which can be attributed to the measurement setup's inherent accuracy errors.
The dashed line represents the case where the experimental and simulation results align.
}
\end{figure}

Figure \ref{fig:sim_sample} compares the experimental results of $j_{{\rm c}}$ with
the calculation results obtained from the 1DM simulation.
Results of the DW velocity as a function of current density
are shown in Appendix \ref{sec:ODM}.
The direction of DW motion agrees between the experiments and simulations,
both with and without PHE correction.
There is a positive correlation between the experimental and simulated values,
and their magnitudes are of the same order. 
These results support the validity of our model,
which includes not only SOT and DMI but also STT and $H_{\rm dep}$.
The inclusion of the PHE correction in the calculation has an impact on the estimated value of the effective magnetic field associated with SOT as described in Eq. (D.2).
When ${\rm PHE}/{\rm AHE}\neq 0$,
the absolute values of the Slonczewski torque efficiency and the SHA become larger compared to the case of ${\rm PHE}=0$. This leads to a stronger driving force for the SOT and a lower simulated value for $|j_{\rm c}|$.

\section{Motion of quasi-Bloch domain wall}

To clarify the behavior of quasi-Bloch DW motion in the presence of SOT,
we present the calculation results of CIDWM in the range of $D=$0-0.2$\,\mathrm{mJ/m}^{2}$ and $\mu_{0}H_{{\rm pin}}=0$-$20\,\mathrm{mT}$.
For the other parameters, we use the fixed values shown in Table \ref{tab:table2}, which are closed to the obtained values for the samples in this study.

\begin{table}[htbp]
\centering
\begin{tabular}{cccccccc}
\hline
$t_{{\rm CFB}}$ &
$M_{{\rm s}}$ &
$\lambda$ & $\Theta_{{\rm SH}}$&
$\beta_{{\rm SO}}$&
$P$ &
$\alpha$ &
${Q}$\\
\hline
$1\,\mathrm{nm}$ & $1.5\,\mathrm{MA/m}$ & $7\,\mathrm{nm}$ & -0.2 & 0.4 & 0.5 & 0.03 & +1\\
\hline
\end{tabular}
\caption{\label{tab:table2}
Fixed parameters in model analysis.}
\end{table}

\begin{figure}[htbp]
\centering
\includegraphics[scale=0.5]{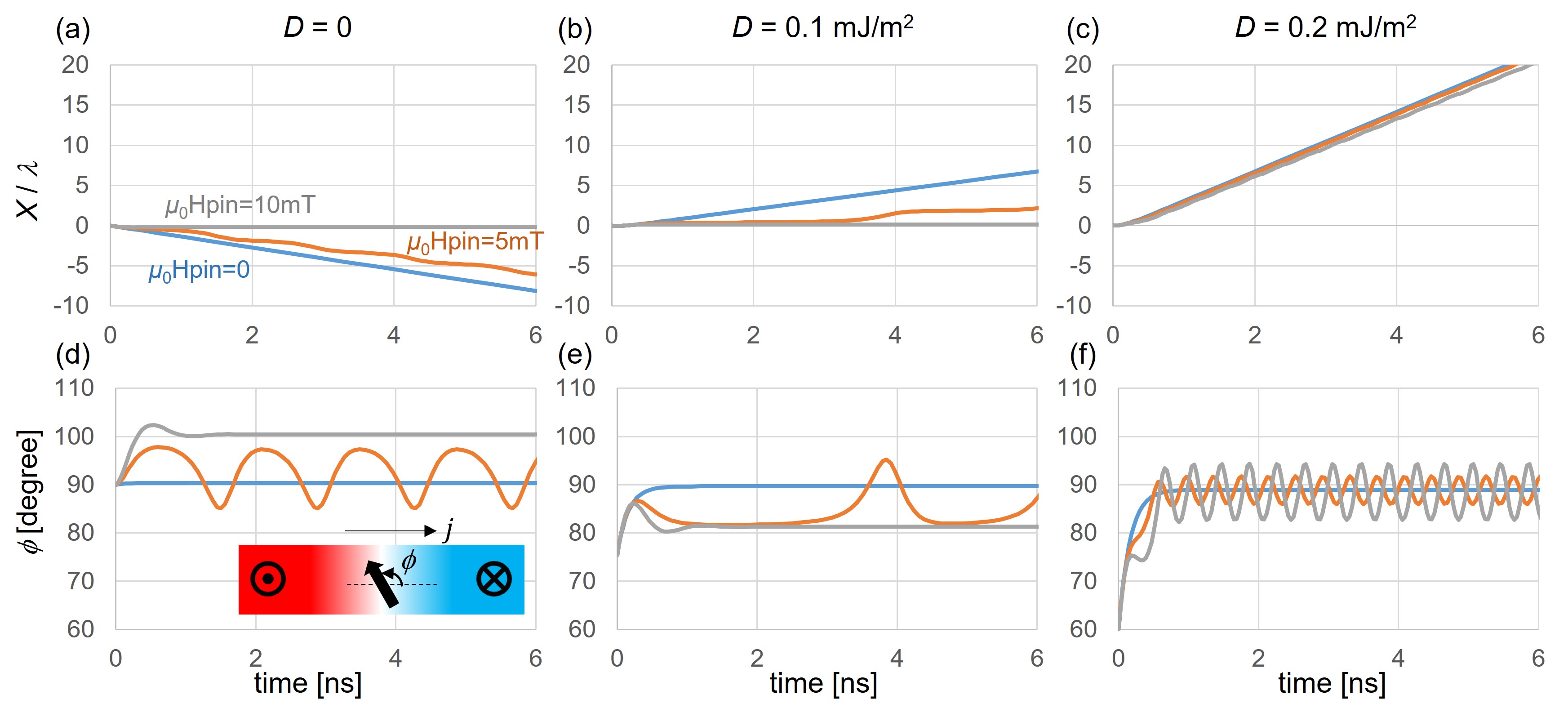}
\caption{\protect\label{fig:dynamics}
(a)-(c) Transient of DW position normalized by the width of DW.
(d)-(f) Transient of DW magnetization angle.
The current density, $j=50\times10^{10}\,\mathrm{A/m}^2$ is applied.}
\end{figure}

\begin{figure}[htbp]
\centering
\includegraphics[scale=0.45]{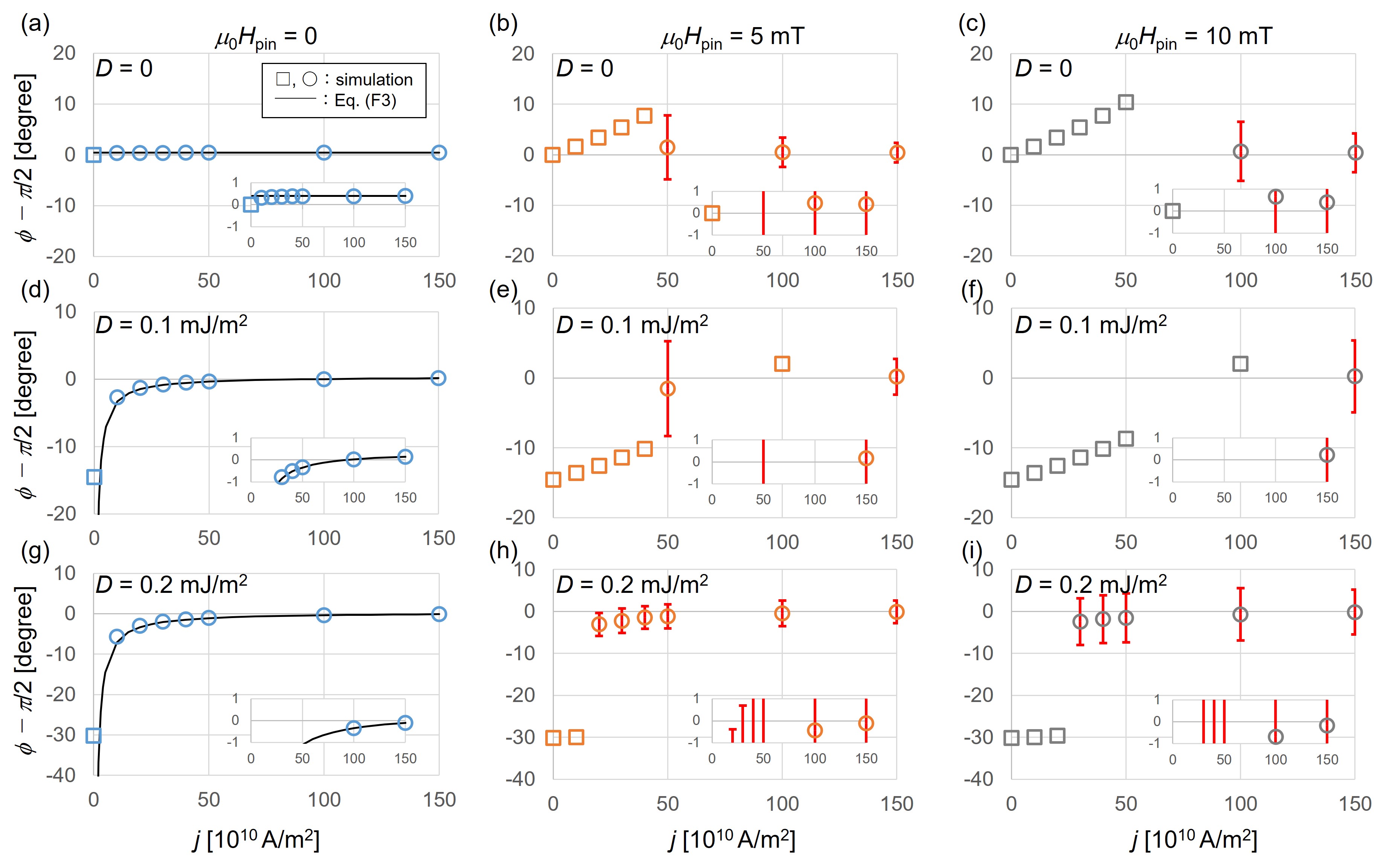}
\caption{\protect\label{fig:angle}
Magnetization angle of DW as a function of current density.
Circle (square) symbols represent the results of $v\neq(=)\,0$. 
The red bars indicate the amplitude of oscillations. The inset provides an enlarged view of the area around $\phi \sim \pi/2$.}
\end{figure}

As an example of the DW dynamics, 
Fig. \ref{fig:dynamics} shows the calculation results for $j=50\times10^{10}\,\mathrm{A/m}^2$.
The DW moves in the direction of current flow for $D\ge0.1\,\mathrm{mJ/m}^{2}$,
but in the direction of electron flow  for $D=0$.
Except for $D=0.2\,\mathrm{mJ/m}^{2}$,
the DW can stop halfway in the case of strong extrinsic DW pinning ($\mu_0 H_{\rm pin}=10\,\mathrm{mT}$).
The angle, $\phi$ to approach $90^\circ$ during current flow is attributed to the influence of the Slonczewski torque, which acts to align $\phi$ toward $90^\circ$. From Fig. \ref{fig:dynamics}, it may appear that $\phi$ is exactly $90^\circ$ in the absence of extrinsic DW pinning.
However, as shown in Fig. \ref{fig:angle}, $\phi$ deviates slightly from $90^\circ$.
This deviation reflects the presence of DMI and STT.
During the DW motion in the presence of extrinsic DW pinning,
the oscillatory behaviors of $X$ and $\phi$ reflect the periodicity of $H_{\rm p}(X)$.
The larger the DW velocity, the shorter the oscillation period.

\begin{figure}[htbp]
\centering
\includegraphics[scale=0.5]{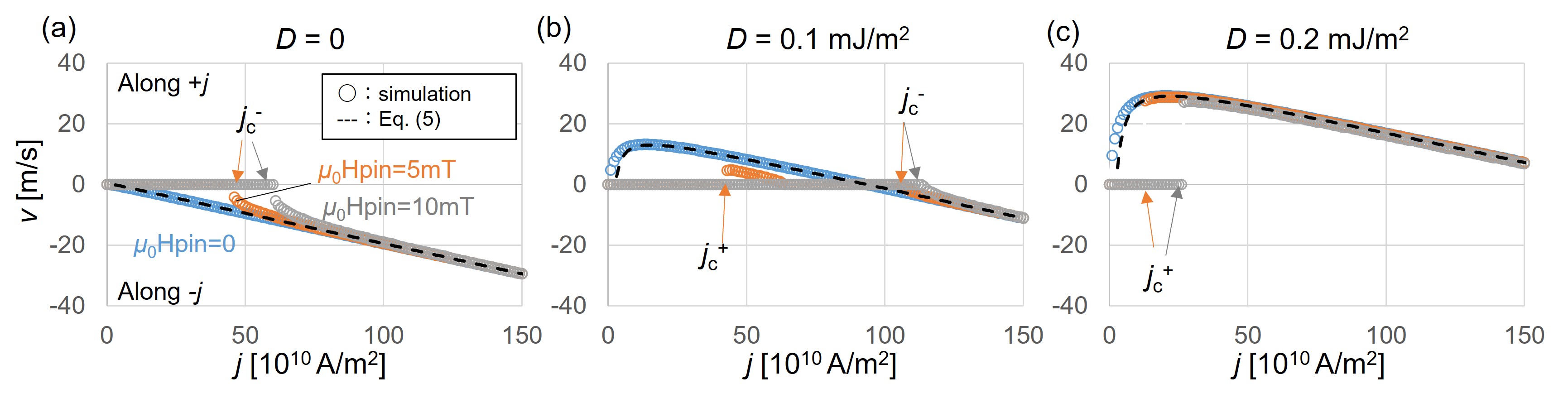}
\caption{\protect\label{fig:v-j_1DM}
DW velocity as a function of current density.}
\end{figure}

Figure \ref{fig:v-j_1DM} shows the DW velocity as a function of the current density ($v$-$j$ curve).
In the case of $D=0$, the DW only moves in the direction of electron flow, and the DW speed increases with increasing $j$.
However, in certain cases with $D>0$, the DW can move in the direction of current flow. 
For example, in the case of $D=0.2\,\mathrm{mJ/m}^{2}$,
the DW speed initially increases with $j$ in regions where $j$ is small,
but after a certain point, the DW speed decreases with $j$.
This behavior is consistent with the previous calculation results \cite{Rizinggard_2017, Kato_2019}.
In the case of $D=0.1\,\mathrm{mJ/m}^{2}$,
the $v$-$j$ curve for $\mu_{0}H_{{\rm pin}}=10\, \mathrm{mT}$ is similar to that for $D=0$,
but the $v$-$j$ curve for $\mu_{0}H_{{\rm pin}}=0\,\mathrm{mT}$ and $5\,\mathrm{mT}$ is similar to that for $D=0.2\,\mathrm{mJ/m}^{2}$.
These results suggest that the direction
of DW motion depends not only on $D$, but also on $H_{{\rm pin}}$ and $j$.

The velocity of the quasi-Bloch DW in the absence of extrinsic DW pinning is given by the following equation, as derived in Appendix \ref{sec:derivation1}:
\begin{equation}
v=\gamma\lambda\left(\frac{\pi}{2}H_{{\rm DMI}}+\xi_{{\rm ST}}j\right)\left(1+\alpha\beta_{\mathrm{SO}}+\alpha\frac{H_{{\rm a}}}{\frac{\pi}{2}\xi_{{\rm SL}}j}\right).\label{eq:v}
\end{equation}
This equation is consistent with the previous study \cite{Torrejon_2014}.
The calculated results using Eq. (\ref{eq:v}) are in good agreement
with the numerical simulation results of Eq. (\ref{eq:1DMeq}) with $H_{\rm pin}=0$, as shown in Fig. \ref{fig:v-j_1DM}.
In Eq. (\ref{eq:v}), the term in the second bracket is approximately
1 due to $\alpha\ll1$, and determinant of the direction of DW
motion is the term in the first bracket. Therefore, in an ideal material
system without extrinsic DW pinning, the relationship between STT and
DMI determines the direction of DW motion.
In our model, $H_{{\rm DMI}}$$\ge0$ and $\xi_{{\rm ST}}j\le0$ are satisfied,
which causes the DW motion in the direction of electron flow due to STT.

It is worth noting that the DW can move in the direction of electron flow even if $D>0$,
as shown in the case  $\mu_0H_{\rm pin}=10\, \mathrm{mT}$ in Fig. \ref{fig:v-j_1DM}(b).
Karnad \textit{et al.} report the observation of CIDWM in the direction of electron flow in
Ta/CoFeB/MgO structures with $\Theta_{\rm SH}<0$ and $D>0$ \cite{Karnad_2018},
which is opposite to the predicted direction
based on the combination of signs for $\Theta_{\rm SH}$ and $D$.
This combination is the same as in our W/CoFeB/MgO structures.
Since a relatively small DMI value ($D\sim0.03\,\mathrm{mJ/m}^2$) is observed in their sample,
it is possible that the crystallinity of the Ta/CoFeB interface in their sample is compromised.
Their results may suggest a similar behavior to that shown in Fig. \ref{fig:v-j_1DM}(b) with $\mu_0H_{\rm pin}=10\,\mathrm{mT}$.
If the DMI can be controlled by adjusting the annealing temperature, it may be feasible to optimize the device's DW velocity according to Eq. (\ref{eq:v}).
By adjusting the DMI filed to counterbalance the effective magnetic field of the STT,
it could be possible to achieve the slower DW speed needed for precise control of the DW position.

Figure \ref{fig:angle} shows the magnetization angle of the DW  as a function of current density.
When the DW does not move ($v=0$), in some cases, 
$\phi$ deviates from the perfect Bloch-state by more than a few degrees.
However, when the DW is in motion ($v\neq0$),
although $\phi$ oscillates, its median value is close to $\pi/2$.
The direction of DW motion is determined by the chirality corresponding to the sign of the  median value of $\phi-\pi/2$.
When the sign of this value is negative (positive),
the right (left)-handed chirality is dominant,
and the DW moves in the direction of current (electron) flow.
It should be noted that even if the DW moves in the direction of electron flow,
the motion is primarily driven by SOT.
Unlike in STT-driven motion, there is no precessional magnetization rotation in the DW. 

\begin{figure}[htbp]
\centering
\includegraphics[scale=0.5]{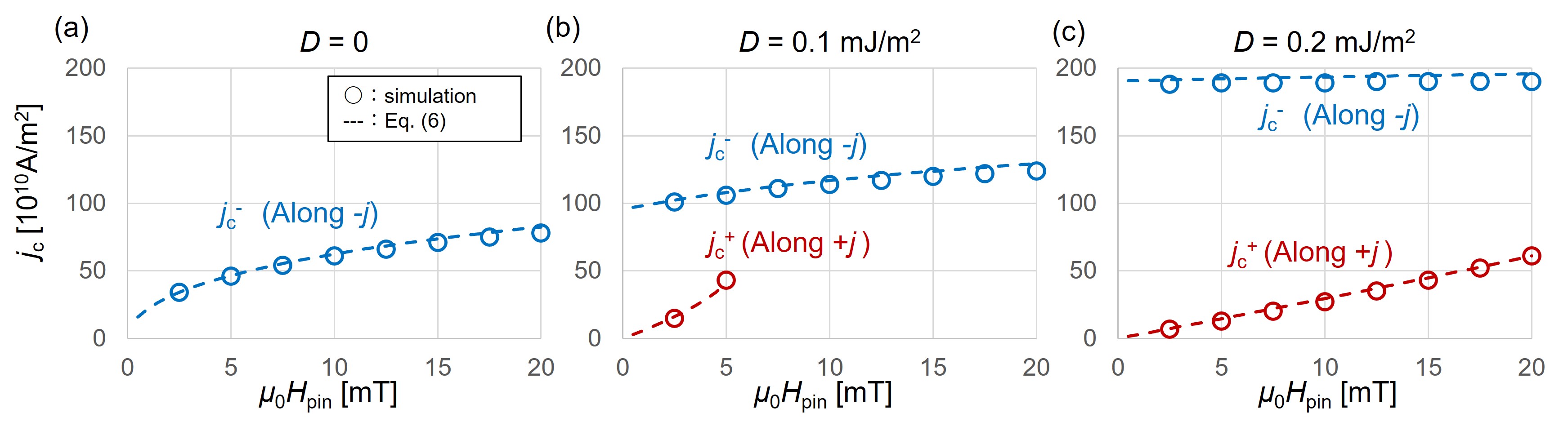}
\caption{\protect\label{fig:jc-Hpin}
Threshold current density for DW motion as a function of extrinsic DW pinning field.
Superscript ``$j_{{\rm c}}$'' indicates the direction of
DW motion ($+$ and $-$ correspond to the current and electron flow,
respectively).}
\end{figure}

Figure \ref{fig:jc-Hpin} shows the threshold current density
for DW motion as a function of the extrinsic DW pinning field.
Since the behavior of $j_{\rm c}$ varies depending on the three DMIs ($D=0,\,0.1,\,0.2\,\mathrm{mJ/m}^{2}$),
we can classify the regions around these values as weak, intermediate, and strong, respectively.
With a weak DMI ($D\sim 0$), the DW motion is in the direction of electron flow.
With a strong DMI ($D\sim 0.2\,\mathrm{mJ/m}^{2}$),
the DW motion is in the direction of current flow in the low-current density region and in the direction of electron flow in the high-current density region.
In these cases, the minimum current density required for DW motion continuously increases with increasing $H_{{\rm pin}}$
in the range of $0\le j \le 200\times10^{10}\,\mathrm{A/m}^2$.
However, at an intermediate DMI ($D\sim 0.1\,\mathrm{mJ/m}^{2}$),  the minimum current density required for DW motion increases discontinuously at a certain point, which corresponds to the condition
for switching direction of DW motion.
A similar discontinuity is also observed in Fig. \ref{fig:jc-Hpin}(c) for $\mu_{0}H_\mathrm{pin} > 20\,\mathrm{mT}$, which is outside the range depicted in this graph.
Therefore, the behavior of $j_\mathrm{c}$ for Fig. \ref{fig:jc-Hpin}(b) and (c) are qualitatively identical.
With weak extrinsic DW pinning, the DW can move in the direction of current flow,
but with strong extrinsic DW pinning, it can only move in the direction of electron flow.
If we were able to prepare samples with similar DMI but different $H_\mathrm{pin}$ values, conducting CIDWM experiments could demonstrate that the direction of the DW and the minimum current density required for DW motion are sensitive to $H_\mathrm{pin}$ in the vicinity of the discontinuity.

The threshold current density of SOT-driven quasi-Bloch DW is given as follows, as derived in Appendix \ref{sec:derivation2}:
\begin{align}
j_{{\rm c}}^{\pm}=&\frac{1}{\pi\left|\xi_{{\rm SL}}\right|\left|\xi_{{\rm ST}}\right|}\Biggl(
\frac{\pi}{2}H_{{\rm DMI}}\frac{\pi}{2}\left|\xi_{{\rm SL}}\right|\mp H_{{\rm pin}}\frac{\pi}{2}\xi_{{\rm FL}} \nonumber \\
&\mp\sqrt{\left(\frac{\pi}{2}H_{{\rm DMI}}\frac{\pi}{2}\left|\xi_{{\rm SL}}\right|\mp H_{{\rm pin}}\frac{\pi}{2}\xi_{{\rm FL}}\right)^{2}\mp2\pi\left|\xi_{{\rm SL}}\right|\left|\xi_{{\rm ST}}\right|H_{{\rm pin}}H_{{\rm a}}}\Biggr),\label{eq:jc_analytical}
\end{align}
where superscript ``$j_{{\rm c}}$'' indicates the direction of
DW motion ($+$ and $-$ correspond to the current and electron flow,
respectively).
The calculated results using Eq. (\ref{eq:jc_analytical}) are in good agreement
with the numerical simulation results of Eq. (\ref{eq:1DMeq}) as shown in Fig. \ref{fig:jc-Hpin}.
The observed discontinuity at $D=0.1\,\mathrm{mJ/m}^{2}$
corresponds to the condition where the root term of Eq. (\ref{eq:jc_analytical})
is zero. Furthermore, the condition for DW motion in the direction
of current flow corresponds to the case where inside of the root
of $j_{{\rm c}}^{+}$ in Eq. (\ref{eq:jc_analytical}) is positive.
In this case, both $j_{c}^+$ and $j_{c}^-$ exist in one structure, as shown in Fig. \ref{fig:jc-Hpin}(b) and (c).
When the current density exceeds a threshold value, the stable DW state transitions from a right-handed chiral structure to a left-handed chiral structure, causing the DW to move in the direction of the electron flow.
To the best of our knowledge,
there has been no actual observations of a phenomenon where the direction of DW motion
switches with current intensity, as shown in this calculation.
This may due to the fact that the nucleation of magnetic domains by SOT magnetization reversal
inhibits the observation of CIDWM in the high-current density region.

\section{Summary}

In this study, we investigate the characteristics of  SOT-driven motion of quasi-Bloch DWs
in perpendicularly magnetized W/CoFeB/MgO ultra-thin films.
Our one-dimensional model, incorporating parameters based on the obtained values for the samples,
successfully reproduces the experimental results of the direction and threshold current density of DW motion, which exhibit variations among different samples.
Our theoretical analysis reveals that during CIDWM, the DW remains in quasi-Bloch-states
and the primary driving force is SOT, rather than STT.
However, STT significantly influences the direction of DW motion.
Additionally, the strength of DMI and extrinsic DW pinning also affect the direction of quasi-Bloch DW motion, in contrast to N\'{e}el-type DW motion.
These findings provide an explanation for the discrepancy between  the experimental direction of DW motion and
the expected direction based on the sign combination of DMI and spin Hall angle.
Furthermore, our study derives analytical expressions for DW velocity and threshold current density, which can be used to easily predict the characteristics of SOT-driven quasi-Bloch DW motion. These analytical equations will be valuable for designing film structures relevant to DW devices, as they allow for the optimization of low currents and appropriate speeds of DW motion required for future devices.


\appendix

\section{\label{sec:VSM}
Vibrating sample magnetometry measurements}

\begin{figure}[htbp]
\centering
\includegraphics[scale=0.65]{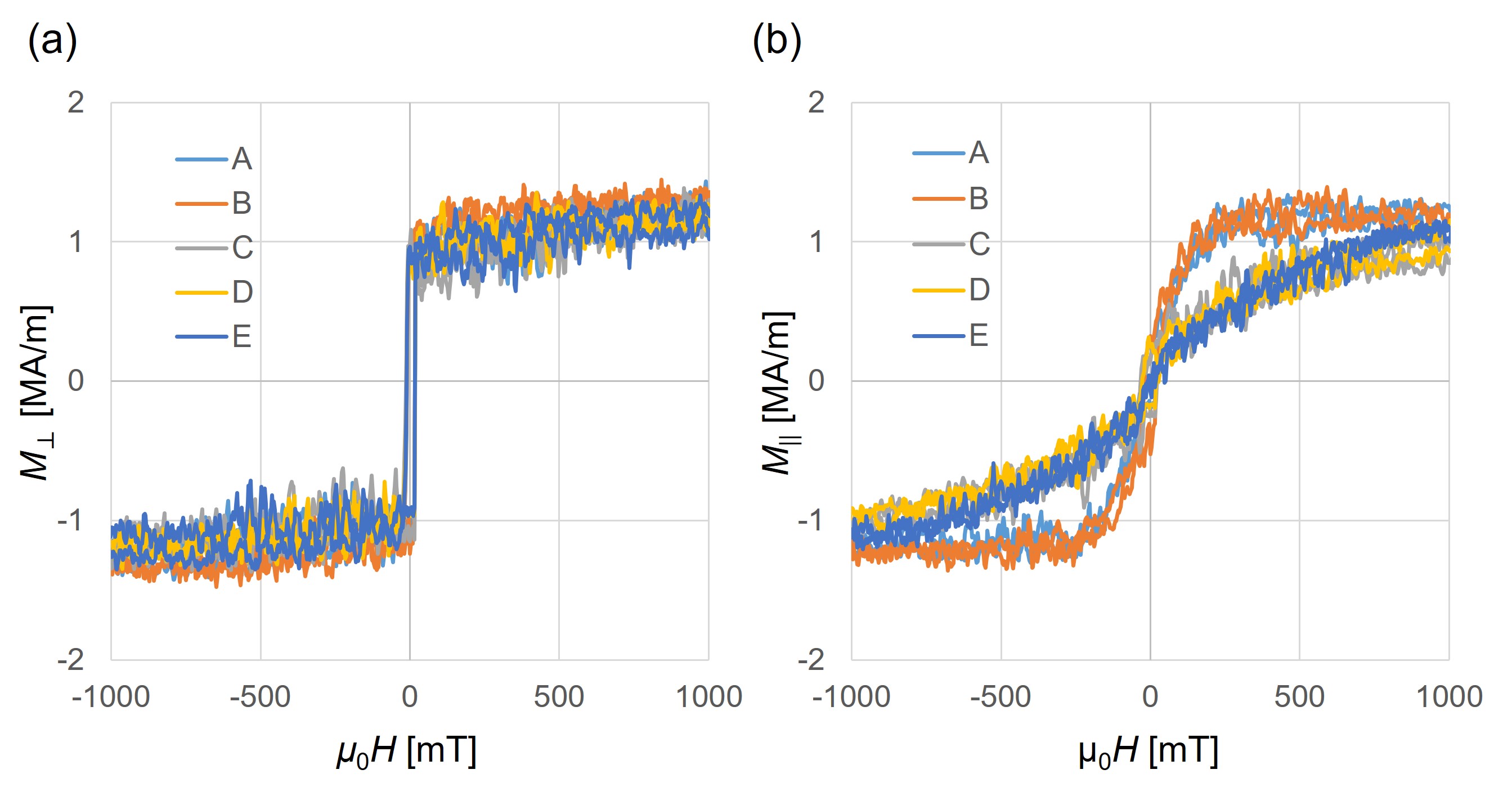}
\caption{\protect\label{fig:VSM}
Magnetization curves measured by VSM. (a) Out-of-plane field. (b) In-plane field. The labels (A-E) correspond to Table \ref{tab:table1}.}
\end{figure}

\begin{figure}[htbp]
\centering
\includegraphics[scale=0.65]{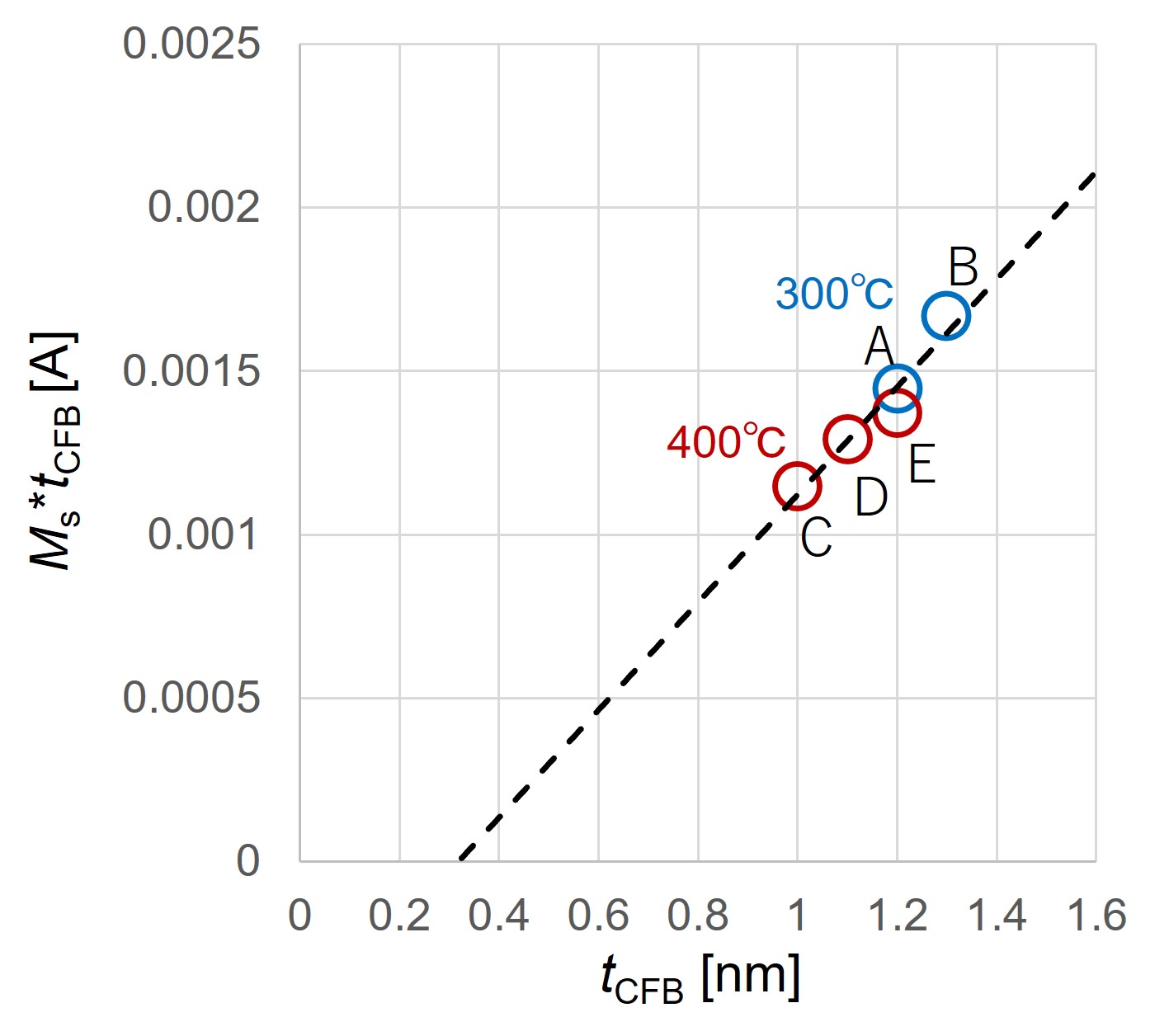}
\caption{\protect\label{fig:Mst}
Products of $M_{\rm s}$ and CoFeB thickness, $t_{\rm CFB}$ as a function of $t_{\rm CFB}$. The labels (A-E) correspond to Table \ref{tab:table1}.}
\end{figure}

We measured the saturation magnetization, $M_{\rm s}$ and the effective PMA,
$K_{\rm eff}$ by using vibrating sample magnetometry (VSM).
Figure \ref{fig:VSM} shows the magnetization curves for each sample,
and Fig. \ref{fig:Mst} shows the products of $M_{\rm s}$ and CoFeB thickness, $t_{\rm CFB}$ as a function of $t_{\rm CFB}$.
By neglecting the annealing temperature dependence of the magnetic dead layer thickness, $t_{\rm d}$,
we can estimate $t_{\rm d}$ to be approximately $0.3\,$nm from Fig. \ref{fig:Mst}.
The values of magnetic properties shown in Fig. \ref{fig:properties} are evaluated considering the value of $t_{\rm d}$.

\section{\label{sec:Depin}
Depinning field measurements}

\begin{figure}[htbp]
\centering
\includegraphics[scale=0.65]{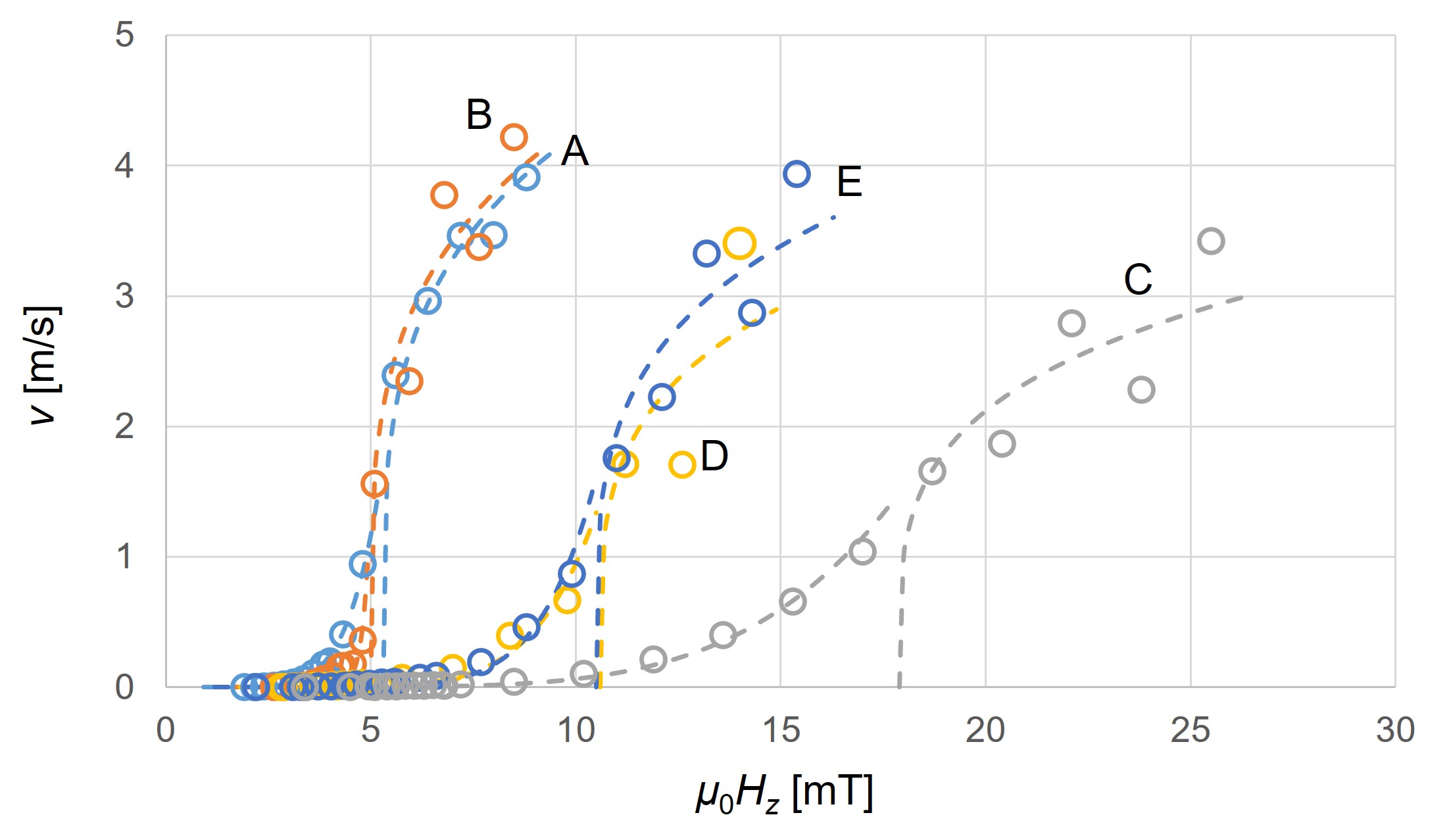}
\caption{\protect\label{fig:v-Hz}
DW velocity as a function of out-of-plane applied field.
The labels (A-E) correspond to Table \ref{tab:table1}. The dotted lines are fitting curves by Eq. (\ref{eq:v_Hz}).}
\end{figure}

We determined the depinning field, $H_{\rm dep}$ by conducting magnetic bubble domain expansion experiments following the methodology described in the previous study \cite{Jagt_2022}.
In these experiments, we measured the DW velocity, $v$ in the presence of an out-of-plane applied field, $H_z$.
The domain velocity at temperature, $T$ can be fitted by the following equations in the creep regime ($H_z < H_{\rm dep}$) and the depinning regime ($H_z > H_{\rm dep}$) :
\begin{equation}
v\left(H_{z}\right)=\begin{cases}
v_{{\rm dep}}\exp\left(-\frac{\Delta E}{k_{{\rm B}}T}\right) & \left(H_{z}<H_{{\rm dep}}\right)\\
\frac{v_{{\rm dep}}}{x_{0}}\left(\frac{T_{{\rm dep}}}{T}\right)^{\psi}\left(\frac{H_{z}-H_{{\rm dep}}}{H_{{\rm dep}}}\right)^{\beta} & \left(H_{z}> H_{{\rm dep}}\right) \label{eq:v_Hz}
\end{cases}
\end{equation}
where $\Delta E=k_{\rm B} T_{\rm dep} [(H_z/H_{\rm dep} )^{-\mu}-1]$ is the creep energy barrier, $k_{\rm B}$ is the Boltzmann constant,
$v_{\rm dep}$ is the DW velocity at $H_z = H_{\rm dep}$,
$T_{\rm dep}$ is the depinning temperature, $\mu=\beta=1/4$ and $\psi= 0.15$ are the critical exponents, and $x_0 = 0.65$ is the universal constant.
Figure \ref{fig:v-Hz} shows $v$ as a function of $H_z$ for each sample.
The value of $H_{\rm dep}$ was obtained by fitting the data using Eq. (\ref{eq:v_Hz})  with $T = 300\,$K assumed.

\section{\label{sec:DMI}
Dzyaloshinskii-Moriya interaction field measurements}

\begin{figure}[htbp]
\centering
\includegraphics[scale=0.4]{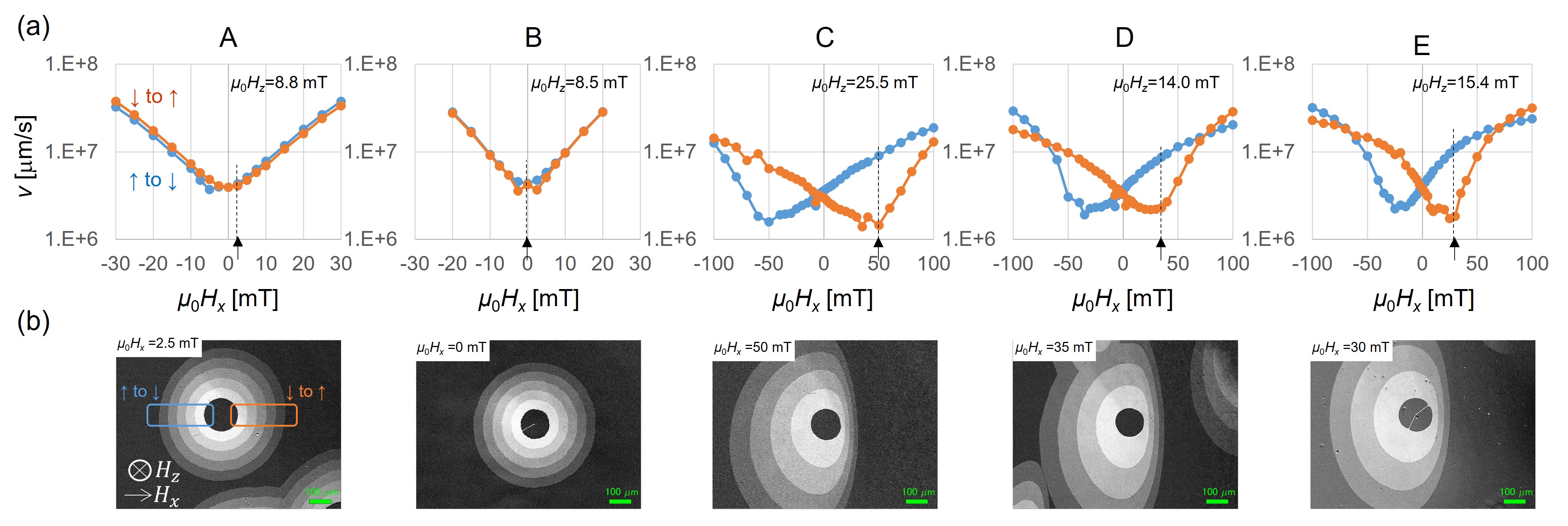}
\caption{\protect\label{fig:v-Hx}
(a) DW velocity as a function of the in-plane applied field.
(b) Image of measured magnetic domain change.
The magnetic domain at $H_x$ indicated by the black arrows in the graph are shown.
The labels (A-E) correspond to Table \ref{tab:table1}.
}
\end{figure}

We conducted magnetic bubble domain expansion experiments to measure the DMI field, $H_{\rm DMI}$ \cite{Rohart_2013, Soucaille_2016, Quinsat_2017}.
In these experiments, the DW velocity, $v$ driven by a pulsed out-of-plane applied field, $H_z$ was measured as a function of the in-plane field, $H_x$.
To ensure that only the influence of $H_{\rm DMI}$ is taken into account in our observations, 
we conducted measurements in the depinning or flow region ($H_z > H_{\rm dep}$) \cite{Dohi_2019}.
Figure \ref{fig:v-Hx} shows the DW velocity as a function of $H_x$ for each sample.
The global minimum of the $v$-$H_x$ curve is defined as $H_{\rm DMI}$. 

\section{\label{sec:Harmonic}
Harmonic Hall voltage measurements}

\begin{figure}[htbp]
\centering
\includegraphics[scale=0.65]{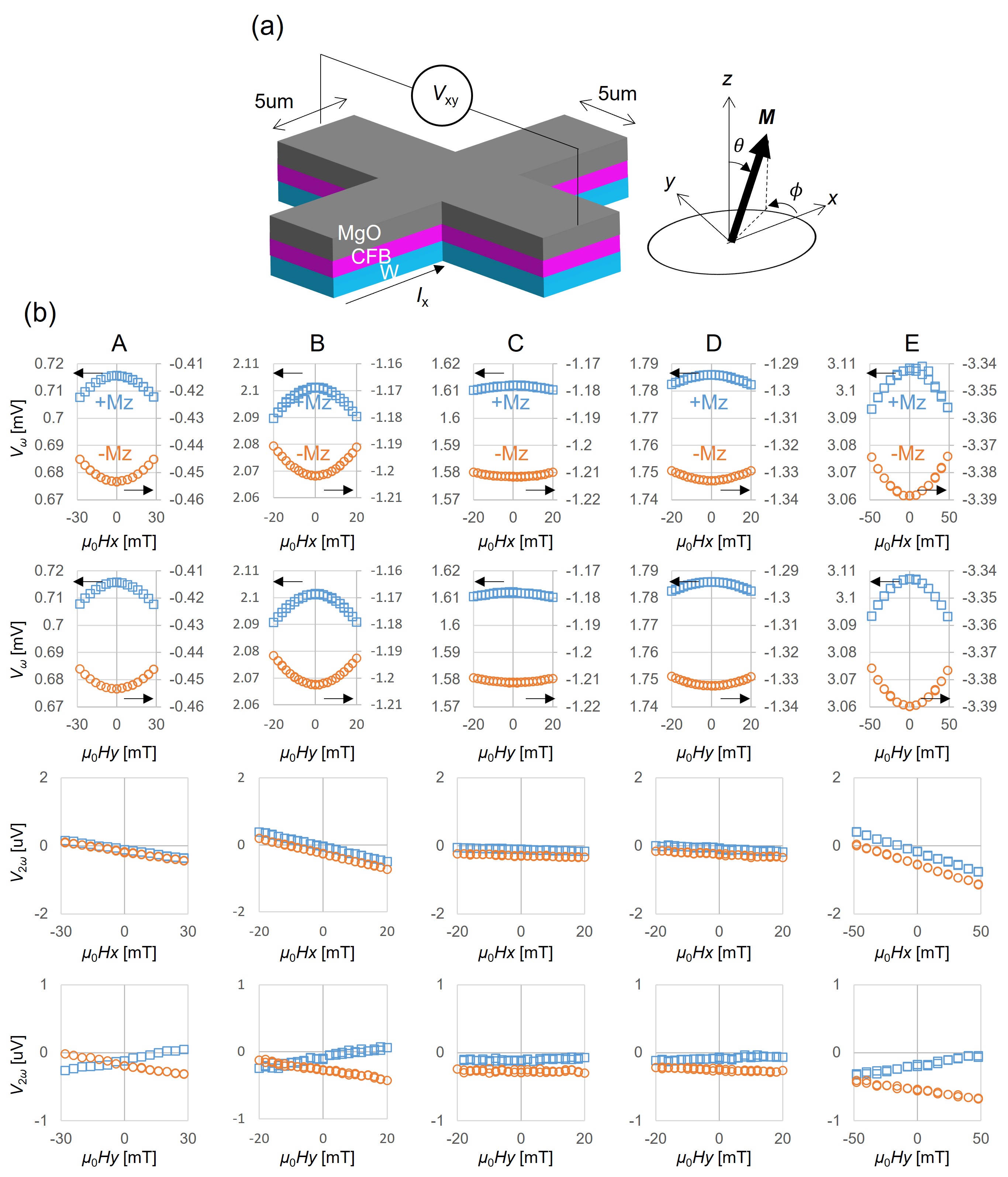}
\caption{\protect\label{fig:Harmonic}
(a) Schematic illustration of setup for harmonic Hall voltage measurements and coordinate system.
(b) Harmonic Hall voltage versus in-plane field. The labels (A-E) correspond to Table \ref{tab:table1}.}
\end{figure}

We evaluated the Slonczewski (SL) and field-like (FL) torque efficiencies using harmonic Hall voltage measurements, following the conventional methodology \cite{Hayashi_2014}.
The frequency, $\omega(2\omega)$ Hall voltage, $V_{\omega(2\omega)}$ was measured under an in-plane field,
$H_{x(y)}$ using the experimental setup depicted in Fig. \ref{fig:Harmonic}(a).
The experimental results of the harmonic Hall voltage as a function of $H_{x(y)}$ for each sample are shown in Fig. \ref{fig:Harmonic}(b).
The effective fields for SL and FL torque are denoted as
\begin{equation}
H_{{\rm SL} \left({\rm FL}\right)}=-2\frac{\partial V_{2\omega}/\partial H_{x\left( y\right)} }{\partial^{2}V_{\omega}/\partial H_{x\left( y\right)}^{2}},
\end{equation}
and SL (FL) torque efficiencies are experimentally defined by $\xi_{\rm SL(FL)}=\mu_0 H_{\rm SL(FL)}/j$,
where $j$ is the current density.
Since the specific resistivities of the CFB and W layers are expected to be roughly similar \cite{Torrejon_2014},
we assume that the same current density occurs in both layers. 
When calculating $\xi_{\rm SL(FL)}$, it is possible to consider a correction for the planar Hall effect (PHE) using the following equation:
\begin{equation}
H_{{\rm SL}\left({\rm FL}\right)}^{{\rm c}}=\frac{H_{{\rm SL}\left({\rm FL}\right)}\pm 2\eta H_{{\rm SL}\left({\rm FL}\right)}}{1-4\eta^{2}}\label{eq:H_SOT_corrected}
\end{equation}
where $\eta=\Delta R_{\rm PHE}/\Delta R_{\rm AHE}$ is the ratio of resistivity of PHE to the anomalous Hall effect (AHE).
While some researchers suggest that the PHE correction may not be necessary \cite{Zhu_2019},
this study investigates both cases where the correction is considered and not considered.
In the case where the PHE correction is considered, a value of $\eta= 0.3$ is used \cite{Skowronski_2016}.

\section{\label{sec:ODM}
One dimensional model simulation of domain wall velocity}

\begin{figure}[htbp]
\centering
\includegraphics[scale=0.65]{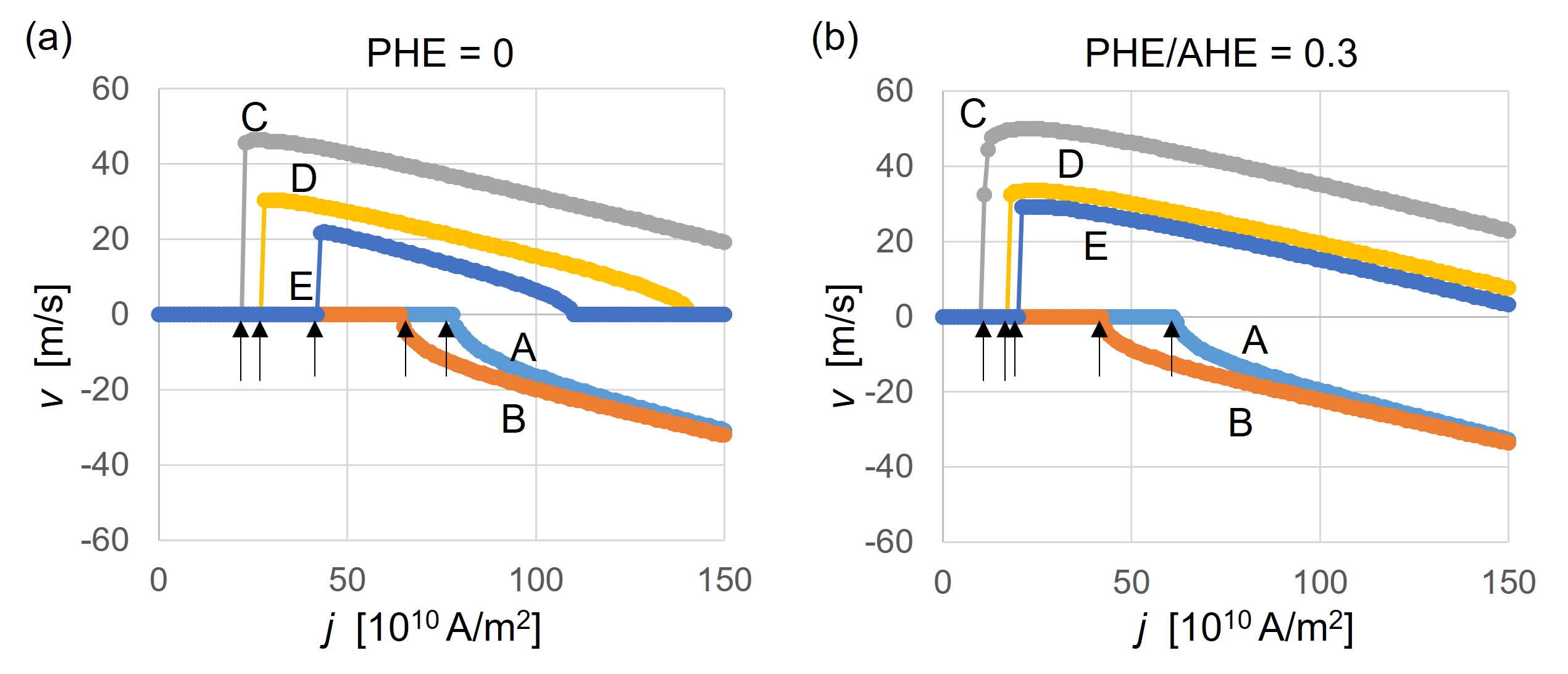}
\caption{\protect\label{fig:v-j_samples}
DW velocity as a function of current density for (a) without and (b) with PHE correction.
The labels (A-E) correspond to Table \ref{tab:table1}.
Arrows represent the threshold current density, $j_{\rm c}$.}
\end{figure}

Figure \ref{fig:v-j_samples} shows the DW velocity, $v$ as a function of current density, $j$ for each sample.
The threshold current density of DW motion for each sample is shown in Fig. \ref{fig:sim_sample}.

\section{\label{sec:derivation1}
Velocity of domain wall motion in the absence of extrinsic pinning}

When the DW reaches an equilibrium state in the presence of currents
(${\rm d}\phi/{\rm d}t=0$), the following equation is satisfied based
on Eq. (\ref{eq:1DMeq}):
\begin{equation}
a\cos\phi+b\sin\phi-\frac{1}{2}c\sin2\phi-d=0,\label{eq:dot-phi}
\end{equation}
where $a=\frac{\pi}{2}\left(\xi_{{\rm SL}}-\alpha\xi_{{\rm FL}}\right)j$,
$b={Q}\alpha\frac{\pi}{2}H_{{\rm DMI}}$, $c=\alpha H_{{\rm a}}$,
and $d=-{Q}\alpha\xi_{{\rm ST}}j$. The stable solution of
Eq. (\ref{eq:dot-phi}) satisfies
\begin{equation}
a\sin\phi-b\cos\phi+c\cos2\phi<0.\label{eq:stable-condition}
\end{equation}
If the solution to Eq. (\ref{eq:dot-phi}) is given by $\phi=\pm\pi/2+\varepsilon$,
Eq. (\ref{eq:stable-condition}) can be approximately written as $\pm\xi_{{\rm SL}}j<0$ when $\alpha\ll1$,
where $\varepsilon$ is the deviation magnetization angle $(|\varepsilon |\ll 1)$ from the perfect Bloch-state ($\phi=\pi/2$).
In the following, we consider the case $j>0$, which leads to $\phi=\pi/2+\varepsilon$
due to $\Theta_{{\rm SH}}<0$ in our model.
By substituting $\phi=\pi/2+\varepsilon$ into Eq. (\ref{eq:dot-phi}),
we can express 
\begin{equation}
\varepsilon\sim\frac{d-b}{c-a}\sim{Q}\alpha\frac{\frac{\pi}{2}H_{{\rm DMI}}+\xi_{{\rm ST}}j}{\frac{\pi}{2}\xi_{{\rm SL}}j},\label{eq:epsilon_1}
\end{equation}
when $\alpha\ll1$. 
As shown in Fig. \ref{fig:angle},
the calculation results using Eq. (\ref{eq:epsilon_1}) are in good
agreement with the numerical simulation results of Eq. (\ref{eq:1DMeq}).
By substituting Eq. (\ref{eq:epsilon_1})
into Eq. (\ref{eq:1DMeq}), we obtain Eq. (\ref{eq:v}).

\section{\label{sec:derivation2}Threshold current density required for domain wall motion}

When DW does not move (${\rm d}\phi/{\rm d}t={\rm d}X/{\rm d}t=0$),
the following equations are satisfied based on Eq. (\ref{eq:1DMeq}):
\begin{align}
a\cos\phi+b\sin\phi-\frac{1}{2}c\sin2\phi-d+{Q}H_{{\rm p}}\left(X\right) & =0,\label{eq:abcd-w.pin}\\
e\cos\phi-b\sin\phi+\frac{1}{2}c\sin2\phi+d+{Q}\alpha^{2}H_{{\rm p}}\left(X\right) & =0,\label{eq:ebcd-w.pin}
\end{align}
where $a=\frac{\pi}{2}\left(\xi_{{\rm SL}}-\alpha\xi_{{\rm FL}}\right)j$,
$b={Q}\alpha\frac{\pi}{2}H_{{\rm DMI}}$, $c=\alpha H_{{\rm a}}$,
$d=-{Q}\alpha\xi_{{\rm ST}}j$, and $e=\alpha\frac{\pi}{2}\left(\alpha\xi_{{\rm SL}}+\xi_{{\rm FL}}\right)j$.
When $\alpha\ll1$, we obtain
\begin{equation}
H_{{\rm p}}\left(X\right)\sim-{Q}\frac{\pi}{2}\xi_{{\rm SL}}j\cos\phi,\label{eq:Hp}
\end{equation}
and 
\begin{equation}
f\cos\phi-b\sin\phi+\frac{1}{2}c\sin2\phi+d\sim0,\label{eq:with-pin_stable}
\end{equation}
from Eq. (\ref{eq:abcd-w.pin}) and (\ref{eq:ebcd-w.pin}), where
$f=\alpha\frac{\pi}{2}\xi_{{\rm FL}}j$. By substituting $\phi=\pi/2+\varepsilon$
into Eq. (\ref{eq:with-pin_stable}), we can express $\varepsilon$
as
\begin{equation}
\varepsilon\sim\frac{d-b}{c+f}=-{Q}\frac{\frac{\pi}{2}H_{{\rm DMI}}+\xi_{{\rm ST}}j}{H_{{\rm a}}+\frac{\pi}{2}\xi_{{\rm FL}}j}, \label{eq:epsilon_2}
\end{equation}
where $\varepsilon$ is the deviation magnetization angle $(|\varepsilon |\ll 1)$ from the perfect Bloch-state ($\phi=\pi/2$).
From Eq. (\ref{eq:Hp}) and
(\ref{eq:epsilon_2}), we obtain 
\begin{equation}
\frac{\pi}{2}\xi_{{\rm SL}}\xi_{{\rm ST}}j^{2}
+\left(\frac{\pi}{2}H_{{\rm DMI}}\frac{\pi}{2}\xi_{{\rm SL}}+H_{{\rm p}}\left(X\right)\frac{\pi}{2}\xi_{{\rm FL}}\right)j 
+H_{{\rm p}}\left(X\right)H_{{\rm a}}=0.\label{eq:j_2}
\end{equation}
Note that $\xi_{{\rm SL}},\xi_{{\rm ST}}<0$, the solution to Eq.
(\ref{eq:epsilon_2}) can be expressed as
\begin{align}
j =& \frac{1}{\pi\left|\xi_{{\rm SL}}\right|\left|\xi_{{\rm ST}}\right|}\Biggl(
\frac{\pi}{2}H_{{\rm DMI}}\frac{\pi}{2}\left|\xi_{{\rm SL}}\right|-H_{{\rm p}}\left(X\right)\frac{\pi}{2}\xi_{{\rm FL}} \nonumber \\
&\pm\sqrt{\left(\frac{\pi}{2}H_{{\rm DMI}}\frac{\pi}{2}\left|\xi_{{\rm SL}}\right|-H_{{\rm p}}\left(X\right)\frac{\pi}{2}\xi_{{\rm FL}}\right)^{2}-2\pi\left|\xi_{{\rm SL}}\right|\left|\xi_{{\rm ST}}\right|H_{{\rm p}}\left(X\right)H_{{\rm a}}} \Biggr)
.\label{eq:j_analytical}
\end{align}
For DW motion in the direction of current and electron flow, we substitute
$H_{{\rm p}}\left(X\right)=+H_{{\rm pin}}$ and $-H_{{\rm pin}}$ into
Eq. (\ref{eq:j_analytical}), respectively. As a result, we finally
obtain Eq. (\ref{eq:jc_analytical}).

\bibliography{reference_v2}

\begin{thebibliography}{59}
\bibitem [{per()}]{periodic}%
  \BibitemOpen
  \href@noop {} {}\bibinfo {note} {The periodic length of extrinsic pinning,
  $L_{{\rm pin}}$ does not significantly affect the velocity and the threshold
  current density of domain wall motion if the order of $L_{{\rm pin}}$ is same
  as the width of the domain wall.}\BibitemShut {Stop}%
\end{thebibliography}

\end{document}